\documentstyle[budapest]{article}
\RequirePackage{rawfonts}
\frompage{000} \topage{000}                                              
\def\rightmark{ Jet Energy Loss in Hot and Dense Matter }  
\def\leftmark{ I. Vitev et al.} 
 
\newcommand{\half}{{\textstyle\frac{1}{2}}}  
\newcommand{\gton}{\mathrel{\lower.5ex \hbox{$\stackrel{> } 
 {\scriptstyle \sim}$}}} 
\newcommand{\lton}{\mathrel{\lower.5ex \hbox{$\stackrel{< } 
 {\scriptstyle \sim}$}}} 
\newcommand{\ee}{\end{equation}} 
\newcommand{\ben}{\begin{enumerate}} \newcommand{\een}{\end{enumerate}} 
\newcommand{\bit}{\begin{itemize}} \newcommand{\eit}{\end{itemize}} 
\newcommand{\bc}{\begin{center}} \newcommand{\ec}{\end{center}} 
\newcommand{\bea}{\begin{eqnarray}} \newcommand{\eea}{\end{eqnarray}} 
\newcommand{\beqar}{\begin{eqnarray}} \newcommand{\eeqar}[1]{\label{#1} 
\end{eqnarray}}  
 
\newcommand{\htR}{\hat{R}}
\newcommand{\htD}{\hat{D}}
\newcommand{\htV}{\hat{V}}

\title{ Jet Energy Loss in Hot and Dense Matter }  
\authors{ 
{I. Vitev$^1$, M. Gyulassy$^1$, and P. Levai$^{2}$ %
}\\[2.812mm] 
{\normalsize 
\hspace*{-8pt}$^1$ Department of Physics, Columbia University, 538 West 
 120-th Street,\\ New York, 
  NY 10027, USA\\[0.2ex]  
\hspace*{-8pt}$^2$  KFKI Research Institute for Particle and Nuclear 
  Physics, \\ P.O. Box 49, Budapest 1525, Hungary 
}}

\abstract{ We discuss the the GLV    
Reaction Operator formalism for computing the non-abelian energy loss 
of jets propagating through hot and dense matter. We incorporate our  
results on induced gluon bremsstrahlung together with the  
effects of nuclear shadowing and multiple scattering in a 
two component soft+hard description of heavy ion reactions at RHIC  
energies. We demonstrate that good agreement between data and theory  
can be achieved in the measured moderate $p_T \leq 5$~GeV window at 
$\sqrt{s}=130$~AGeV  and  we extend our predictions for the 
high $p_T$  part of the hadronic spectra. 
We focus on the perspectives of using  jet tomographic methods for 
probing not only the density and geometry of the quark-gluon plasma 
but also its parton composition at various center of mass energies. }

  
\begin{document} 
  
\maketitle 
\setcounter{page}{1} 
 
\section{Introduction}\label{intro} 
 
Coherent scattering effects on induced photon/gluon emission 
have been studied in both  
abelian and non-abelian gauge theories. Destructive interference 
of  photon radiation from initially on-shell 
fast electrons propagating and multiply interacting in  
matter was first discussed by Landau and Pomeranchuk~\cite{LAND} and 
independently by Migdal~\cite{MIGD}. Photon bremsstrahlung  
off asymptotically prepared jet states was found to be suppressed in the
infrared ($\omega \rightarrow 0$) region relative to the naive 
Bethe-Heitler limit~\cite{BH} of radiation resulting 
from independent collisions. Those results have been 
experimentally confirmed at SLAC~\cite{SC}.     
 
The non-abelian generalization of Bethe-Heitler radiation was discussed 
by Bertsch and Gunion~\cite{BG} in the presence of a single scattering center. 
In a QED-like scenario the case of multiple interactions was first  
investigated by Gyulassy and Wang~\cite{GW}. The non-abelian nature of 
QCD coherence was later employed by BDMPS~\cite{BDMPS} to show that in the 
high energy regime the mean energy loss of asymptotic jets  is 
proportional to  the square of the length of the medium, i.e. 
$\Delta E \sim L^2$. Those results were obtained for ``thick'' 
plasmas with a very large number of soft scatterings. Discussion of
gluon bremsstrahlung in the path integral formalism can be found in
Refs.~\cite{Z,UW}.       
  
The GLV reaction operator approach~\cite{GLV} to non-abelian parton 
energy loss relies on a systematic expansion  of induced gluon radiation 
associated with jet production in a dense QCD plasma in terms of correlations
between multiple scattering centers. 
Analytic expressions for the induced inclusive gluon transverse
momentum and light-cone momentum distributions are derived to all orders in
powers of the opacity of the medium, $\chi = N\sigma_g/A=L/\lambda_g$.    
The analytic solution to all orders in opacity generalizes
previous continuum results  by allowing for arbitrary correlated
nuclear geometry and evolving screening scales as well as the inclusion of
finite kinematic constraints. In comparison to data  we use numerical results
corrected up to third order in $\chi$ that allow us to extend jet quenching
computations down to parton energies $E\sim 5$~GeV. For gaining theoretical
insight, however, it is useful to resort to analytic formulas that neglect 
some of the kinematic constraints but capture the essential features of
parton energy loss.
The dominant first order energy loss can be written as
\beqar
\Delta E^{(1)} & = & \frac{C_R\alpha_s}{2}\int^\infty_{\tau_0} d \tau\,
\frac{\mu^2(\tau)}{\lambda(\tau)}\;(\tau-\tau_0) \log 
\frac{2 E}{\mu^2 L}\; ,
\eeqar{de11}     
where $C_R$ is the second casimir in the representation of the jet and the
$\mu^2(\tau)/\lambda(\tau)$ is the local ``transport coefficient".
In a hot and dense medium in local thermal equilibrium  $\mu^2(\tau) = 
4\pi \alpha_s T^2(\tau)$  and  perturbatively  one can express 
$\mu^2(\tau)/\lambda(\tau) 
= (C_T C_A/d_A)  4 \pi \alpha_s^2\, \rho(\tau)$, where $C_T$ is the color 
charge of the target and $d_A$ is the dimension of the adjoint 
representation. This formulation allows to account for the realistic 
dynamical  time evolution of the systems created in ultra-relativistic
nucleus-nucleus collisions.

Jet tomography is the QCD analog of conventional X-ray tomography that
exploits the attenuation of high energy jets produced in nuclear 
matter~\cite{Gyulassy:2001zv}. 
We use the term ``jet tomography'', inspired by a seminar by 
Istvan Lovas~\cite{istvan} entitled ``Vector meson tomography of the QGP'', 
to contrast the particular advantages of high $p_T$ PQCD probes 
of the density evolution of partonic matter formed in A+A reactions.
The dependence of the non-abelian energy loss on the density and 
geometry of the medium as seen in Eq.~(\ref{de11}) provides a rigorous 
theoretical framework for the jet tomographic analysis discussed 
in Refs.~\cite{gvw,Levai:2001dc,Fai:2001vz,gvw2,Vitev:2001zn}. 
The experimental discovery of a factor of $\sim 3$ 
suppression of moderate $p_{\rm T}  
\lton 4$~GeV $\pi^0$'s in central $Au+Au$ reactions by  
PHENIX~\cite{Adcox:2001jp}  and the discovery of large transverse 
asymmetries in non-central collisions for $p_{\rm T} \lton 5$~GeV  
by STAR~\cite{starv2} have confirmed that the high $p_T$ frontier at RHIC does
in fact  provide a wide range of new physics opportunities.

We here present the results of our calculation of the suppression of
high $p_T$ particle spectra relative to the binary collision scaled
$pp$ reference baseline and the $p_T$ and centrality dependence of the
$p/\pi$ ratios~\cite{Vitev:2001zn}. We find good agreement  between 
data and theory for jet energy loss driven by initial gluon rapidity 
density $dN^g/dy \simeq 800$. One of the descriptions of soft 
parton production, based on the saturation models~\cite{SATUR}, suggests 
slow variation of the number densities of quanta per unit rapidity 
achieved in the early stages of A+A reactions  with $\sqrt{s}$ 
and therefore a slow  variation of the  quenching factor. 
In contrast, at SPS energies  moderate $p_T$ neutral pions were found 
to be enhanced$^a$ by a factor $\sim 2$~\cite{WANG}. We propose 
that one possible solution to this puzzle may be related to differences in 
the composition of the non-abelian plasma (quarks vs. gluons), an important
detail that has so far been ignored in jet tomographic studies.

\section{ Induced Gluon Radiation }\label{radiate}

The GLV formalism~\cite{GLV} for  multiple elastic and inelastic 
interactions inside nuclear matter expands the  differential probability 
of observing a final state jet or jet+gluon system (described by a set of 
quantum numbers $\{\alpha\}$) in orders of 
the correlations between multiple scattering centers, i.e. $P(\{\alpha\}) 
= \sum_n P^{(n)}(\{\alpha\}) $. In the high 
energy eikonal approximation$^b$ a major simplification occurs  as a result
of  the well defined path ordering of sequential interactions inside 
the medium. It is therefore possible to build all relevant classes 
of amplitudes  by subsequent insertion of single Born or ``direct" ($\htD$)
and double Born or ``virtual" ($\htV$) interactions. At the probability level
the recursion relation is generated by the reaction operator~\cite{GLV} 
$ \htR = \htD^\dagger \htD + \htV^\dagger + \htV $. The virtual corrections 
 $\htV^\dagger + \htV$ to
the naive elastic scattering component  $\htD^\dagger \htD $ ensure  
unitarity in the  GLV formalism. 
The approach described here is quite general, i.e. within the
framework of the approximations stated  for an arbitrary initial condition 
described by an amplitude $A_0$,  $P^{(n)}(\{\alpha\}) \propto A_0^\dagger 
\; ( \htR )^n  \,  A_0$. The technical part of the calculation  includes 
solving for the color and kinematic structure of the the ``direct" and 
``virtual" operators. Recently this was illustrated via a computation 
of the elastic broadening of jets propagating in
nuclear matter~\cite{GLVEL}.

For the case of jets produced at finite time $t_0$ in heavy ion reactions the
solution  for the double differential gluon radiation intensity 
was given in Ref.~\cite{GLV}
\beqar
\frac{dI^{ind.}}{dx\, d^2 {\bf k}} &=& 
\frac{C_R \alpha_s}{\pi^2}  \sum\limits_{n=1}^\infty  \prod_{i=1}^n 
\int\limits_0^{L-\Delta z_{1}  \cdots - \Delta z_{i-1} }  
 \frac{d \Delta z_i}{\lambda_g(i)} \int  \left(d^2{\bf q}_{i}\, 
\left[\bar{v}_i^2({\bf q}_{i}) - \delta^2({\bf q}_{i}) \right]\right) 
\, \times \nonumber \\[1.ex]
&& \times \left( -2\,{\bf C}_{(1, \cdots ,n)} \cdot 
\sum_{m=1}^n {\bf B}_{(m+1, \cdots ,n)(m, \cdots, n)} \, \times \right.
\nonumber \\[1.ex]
&&  \quad \left.  \times \, \left[ \cos \left (
\, \sum_{k=2}^m \omega_{(k,\cdots,n)} \Delta z_k \right)
-   \cos \left (\, \sum_{k=1}^m \omega_{(k,\cdots,n)} \Delta z_k \right)
\right]\; \right), 
\eeqar{ndifdis} 
where  ${\bf C}_{(1, \cdots, n)} = \half \nabla_{{\bf k}} 
\log ({\bf k} - {\bf q}_{1}- \cdots  - {\bf q}_{n})^2$, 
${\bf B}_{(m+1, \cdots ,n)(m, \cdots, n)} \equiv {\bf C}_{(m+1, \cdots, n)}- 
 {\bf C}_{(m, \cdots, n)}$ are the Bertsch-Gunion terms~\cite{BG}, and  
$\omega_{(k,\cdots,n)} = 1/(2xE\,|{\bf C}_{(k, \cdots, n)}|^2 ) $ are the
interference phases that produce the non-abelian analog of the LPM 
effect~\cite{LAND,MIGD}. The destructive interference pattern can be seen at
a {\em differential level} in Eq.~(\ref{ndifdis}). In the calculation of the
gluon bremsstrahlung we use a normalized color-screened Yukawa potential
leading to  
$ \bar{v}_i^2({\bf q}_{i}) = \mu^2/ (\pi({\bf q}_{i}^2+\mu^2)^2)$.  One
unexpected result was that the series in  Eq.~(\ref{ndifdis}) converges
very fast~\cite{GLV,FLUC}. Even more surprisingly, the first term was 
found to give the dominant contribution and by integrating  over the 
transverse momentum  ${\bf k}$ and light-cone momentum fraction  $x$ of 
the gluon one recovers  the result of Eq.~(\ref{de11}).  
In computing the spectrum of gluons and the energy loss of jets we take into 
account numerically corrections up to third order in the opacity $\chi$ and 
the evolution of the density of the system due to the longitudinal Bjorken
expansion.

\section{ Hadronic Spectra at Moderate to High $p_T$ }\label{pqcd}

The computed energy loss spectrum is employed in the calculation of the
quenched spectrum of hadrons. A jet of flavor $c$ and  transverse
momentum $p_c$ produced in a hard PQCD scattering $a+b\rightarrow c+d$
is attenuated prior to hadronization by the radiative 
energy loss to $p_c^* = p_c (1-\epsilon)$. This 
shifts the hadronic fragmentation fraction $z_c=p_h/p_c$ 
to $z_c^* = z_c /(1-\epsilon)$.

The invariant distribution of hadrons  reduced by the energy loss
in  $A+A$ collision at impact parameter ${\bf b}$ is then given by
\begin{eqnarray}
\label{fullaa}
&&E_{h}\frac{dN_{h}^{AA}}{d^3p} = T_{AA}( {\bf b} )
        \sum_{abcd}\!  
        \int\!\!dx_1 dx_2  \;   d^2k_{T,a}d^2k_{T,b} 
        g(\vec{k}_{T,a}) g(\vec{k}_{T,b})
        \nonumber \\
        && \ \ \ \  
f_{a/A}(x_1,Q^2) f_{b/A}(x_2,Q^2)\
             \frac{d\sigma^{ab \rightarrow cd}}{d{\hat t}} 
\int d\epsilon \; P(\epsilon,p_c)
\frac{z^*_c}{z_c}
   \frac{D_{\pi^0/c}(z^*_c,Q^2)}{\pi z_c} \,\,\, ,
\label{pqcdsp}
\end{eqnarray}
where $T_{AA}({\bf b})$ is the Glauber profile density. 
The fragmentation functions $D_{h/c}(z,Q^2)$ is taken 
from BKK~\cite{BKK95} and the structure functions for $f_{a/A}(x,Q^2)$ 
(we take the GRV94 LO~\cite{GRV94}) include the 
isospin dependence. Nuclear shadowing, $k_{\rm T}$ broadening and 
Cronin effect are taken into 
account as  in~\cite{XNWint,PLF00,Eloke}. The modification of the 
fragmentation functions due to gluon radiation can be seen in 
Eq.~(\ref{pqcdsp}) and is also discussed in~\cite{WW}. 
The FFs carry the information of the medium induced jet
energy loss but the observed suppression of high $p_T$ is a result 
of the full calculation that folds in  a variety of nuclear effects as 
well as the shape of the initial jet spectrum.
 
In order to extend our calculations to lower transverse momenta of
hadrons we  discuss a phenomenological soft component motivated by the 
string and baryon junction picture.
We patrametrize  here this component as follows:
\begin{equation}
\frac{dN_s ({\bf b})}{dyd^2{\bf p}_{\rm T}} = 
\sum\limits_{\alpha=\pi,K,p,\cdots}
\frac{dn^\alpha}{dy}({\bf b})\;\frac{e^{-p_{\rm T}/T^\alpha({\bf b})}}
{2\pi (T^{\alpha}({\bf b}))^2}
\; . \label{flavslope}
\end{equation}
Such  seemingly ``thermal" distribution  of hadrons is observed already in
$e^+e^-$ collisions and may be related to Gaussian fluctuations in 
the string tension~\cite{bialas}. 
As in string models the soft  component is assumed 
to scale with the number of participants ($N_{part}$). 
In Eq.~(\ref{flavslope}) we also account for the possibly different 
mean inverse slopes $T^\alpha$ for baryons and mesons. 
In the junction picture~\cite{junction}, 
the large $T^{B}$ may arise from the predicted~\cite{dima96} 
smaller junction trajectory slope $\alpha_J^\prime\approx \alpha_R^\prime/3$.  
This  implies that the {\em effective} string tension
is  three times higher than $1/(2\pi\alpha_R^\prime)\approx 1$~GeV/fm
leading in the massless limit to  $\langle p_{\rm T}^2 \rangle_J \simeq 
3 \, \langle p_{\rm T}^2 \rangle_R$.   
In terms of the string model the factor three enhancement of the
mean square $p_{\rm T}$ is due to the random walk in 
$p_{\rm T}$ arising from the decay of the three strings attached to 
the junction. Naively, we would thus expect
$T^{B} \simeq  \sqrt{3}\, T^{\pi}$ predicting an approximately constant 
$\sqrt{\left\langle p_T^2 \right\rangle^B}$  for all baryon species. 

\begin{center}
\vspace*{8.7cm}
\includegraphics{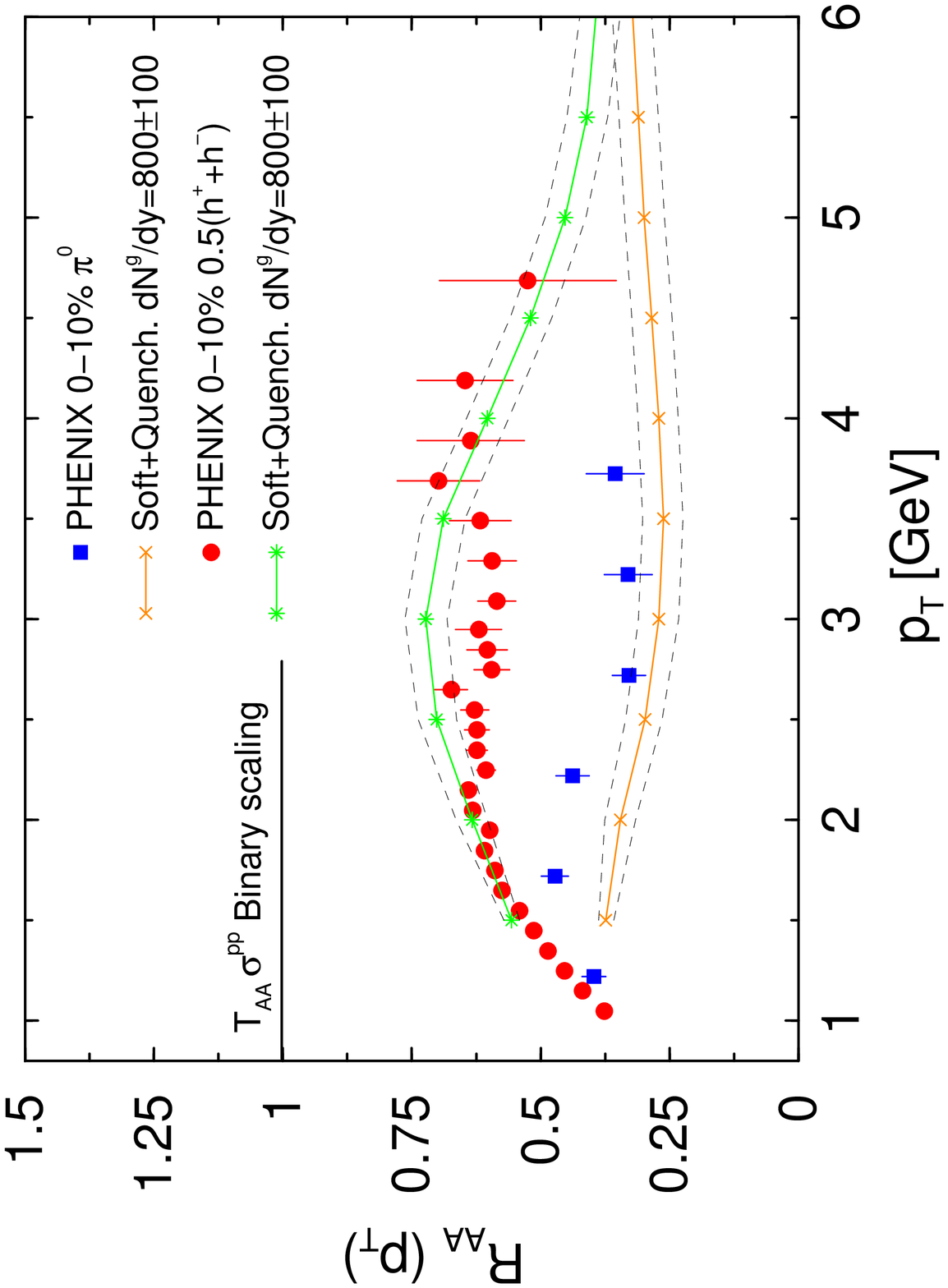}
\vspace{-3.4cm}
\end{center}
\begin{center}
\begin{minipage}[t]{11.5cm}
         {\bf Fig. 1.} {\small  The ratio of charged hadron  and  $\pi^0$ 
multiplicities to the binary collision scaled $\bar{p}p$ result
is shown from~\cite{Adcox:2001jp}.
The curves utilize the GLV quenched hard spectrum and the string and baryon
junction  soft component Eq.~(\ref{flavslope}). }
\end{minipage}
\end{center}

Fig.~1 shows the ratio $R_{AA}$ of the differential particle yields relative 
to the binary collision scaled $\bar{p}p$ for inclusive charged hadrons and 
neutral pions. We obtain reasonable agreement with data if the jet energy loss 
is driven by initial gluon rapidity density $dN^g/dy \simeq 800$. This 
number is lower by $\sim 30\%$ from existing estimates.  The 
difference in the suppression factor of $\pi^0$ and $0.5(h^+ + h^-)$ is
understood through the possibly different baryon and meson production
mechanisms in the moderate high $2 \leq p_T \leq 5$~GeV window.  
In our calculation pion production becomes PQCD dominated for 
$p_T >2$~GeV and correspondingly suppressed by the jet energy loss. In
contrast, baryon production in the region of interest is dominated through the
junction mechanism by baryon transport in rapidity and moderate $p_T$. This
accounts for the different suppression of neutral pions and inclusive charged
hadrons as seen in Fig.~1. At large $p_T$ baryon production also becomes 
perturbative, leading to a common suppression factor. This is manifest in the 
$p_T > 5$~GeV region of Fig.~1.

\begin{center}
\vspace*{8.7cm}
\includegraphics{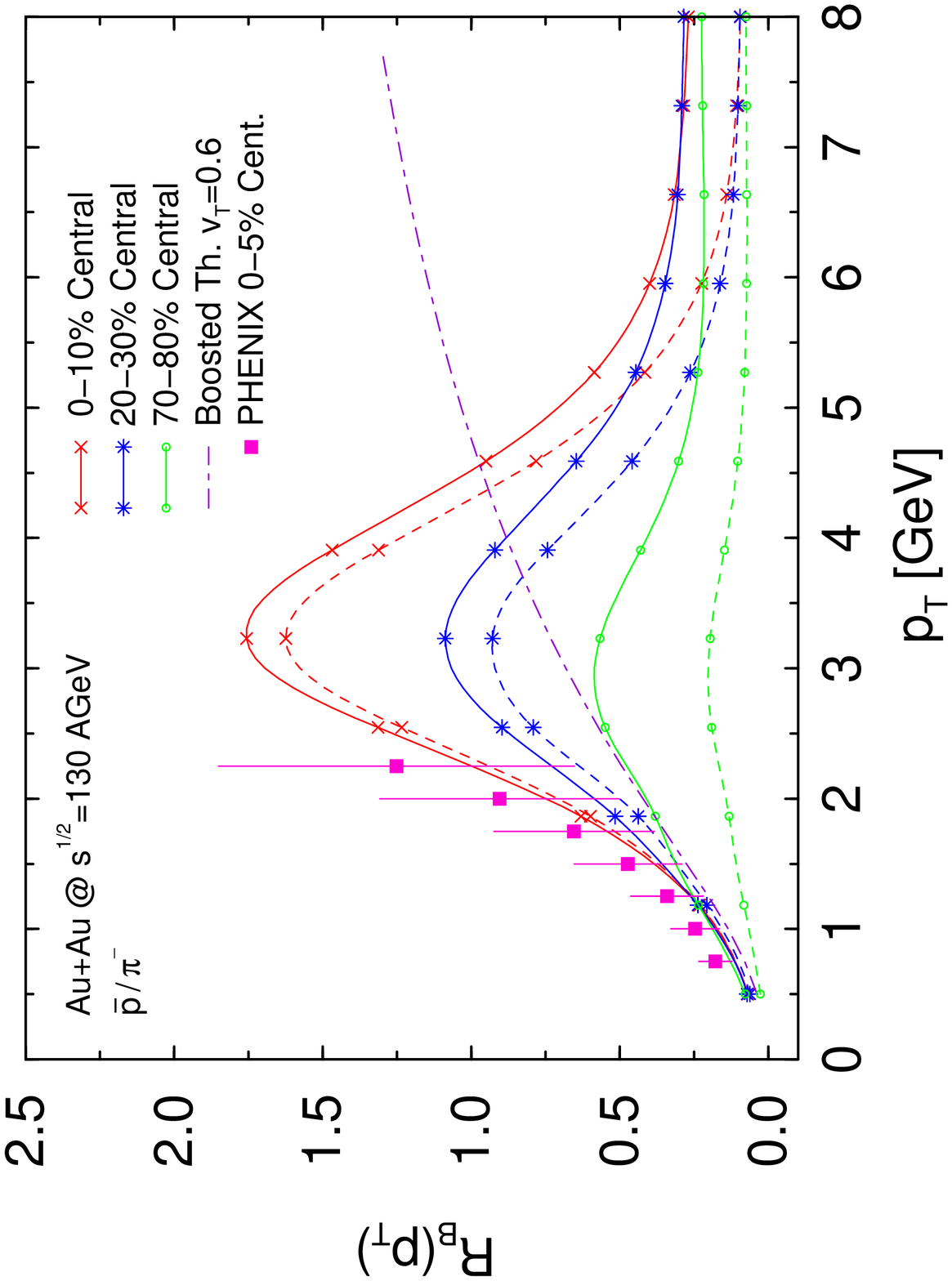}

\includegraphics{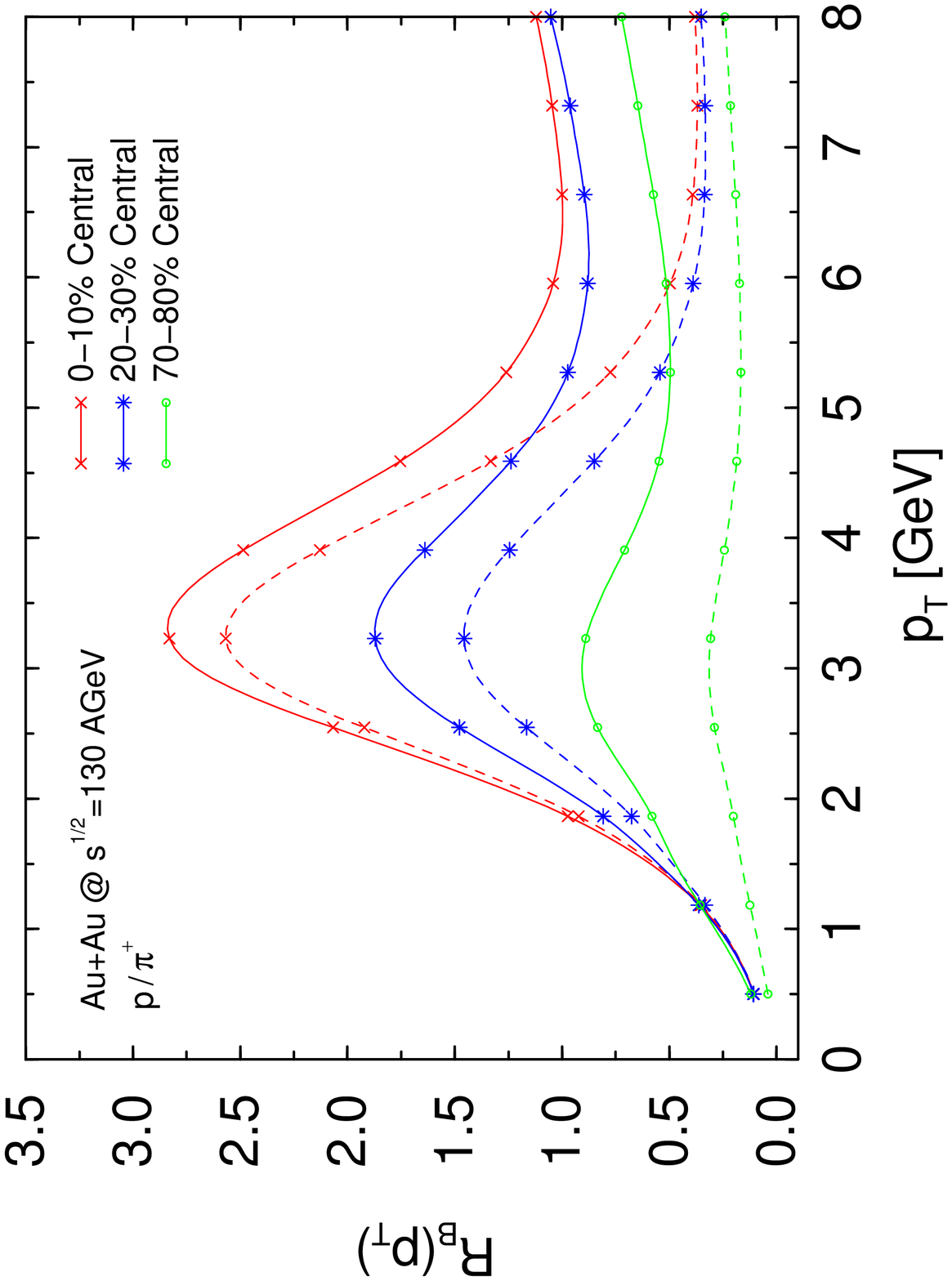}

\vspace{-4.3cm}
\end{center}
\begin{center}
\begin{minipage}[t]{11.5cm}
         {\bf Fig. 2.} {\small  The centrality dependence
of $p/\pi$ is predicted  for three different centralities. 
Solid (dashed) lines correspond to $A^1$ ($A^{4/3}$) scaling of
the baryon junction component.
The ratio of $\bar{p}$  and $\pi^-$ fits to  PHENIX data 
on central reactions is shown for comparison. A 
boosted thermal source (dashed line) is also shown. The left/right panel
reflects negative/positive hadrons.}
\end{minipage}
\end{center}

The interplay between soft and hard physics at RHIC is possibly most 
clearly seen in the $p_T$ differential baryon/meson ($R_B$) ratios.  
Our predictions for the centrality and the $p_T$ dependence of 
$\bar{p}/\pi^-$ and $p/\pi^+$ are  given in Fig.~2. An extended 
discussion can be found in Refs.~\cite{Vitev:2001zn,gvlong}.
In central collisions the interplay  between the anomalous 
baryon component and the quenched PQCD component of 
pions leads  to maximum of $R_B$ near  $p_{\rm T}\sim 3-4$~GeV/c. 
At large  $p_{\rm T} \geq  5-6$~GeV/c we predict   
a gradual decrease of $R_B$ {\em below unity}  consistent  with the 
the PQCD baseline calculations~\cite{Vitev:2001zn}. In Fig.~1  we have 
included through error bands  the factor of $\sim 3$ uncertainty in the 
fragmentation functions into $\bar{p},p$ at high $p_{\rm T}$.
The solid and dashed curves reflect the difference between the 
$N_{part}$ and $N_{part}^{4/3}$ scaling of the junction component.

In peripheral reactions the size of the interaction region 
as well as the initial density of the medium decrease, leading to 
a reduction of energy loss. The absence of  quenching
reduces the observability of the anomalous component and 
the $p/\pi$ ratio may stay  below unity for all $p_{\rm T}$. 
The case of peripheral reactions is hence similar to $\bar{p}p$ collisions.
The experimentally testable  prediction of the model is 
therefore that the maximum of the $R_B= p /\pi $ ratio decreases
with increasing impact  parameter, decreasing participant number, or 
equivalently decreasing  $dN^{ch}/dy$. The reduction of $R_B$ at large 
$p_T$ is also an important prediction of our computation and is not 
seen by other models attempting fits only in regions of $p_T$ where
experimental data already exists.

\section{ Jet Tomography of the QGP Partonic Composition }\label{compo}

Two extreme scenarios for the composition of the quark-gluon plasma have been 
extensively discussed in the literature. In one scenario, based on saturation 
models~\cite{KOVCH}, the hot and dense matter created in A+A collisions
is interpreted as purely gluonic degrees of freedom. Alternatively, quark 
coalescence models~\cite{COAL} 
advocate the creation of dense quark-antiquark matter as a 
result of the  vary fast decay of {\em massive} gluons into quark degrees of
freedom~\cite{UHPL}. We here propose that jet tomography is a powerful tool 
that can distinguish between   different scenarios.  

\begin{center}
\vspace*{8.7cm}
\includegraphics{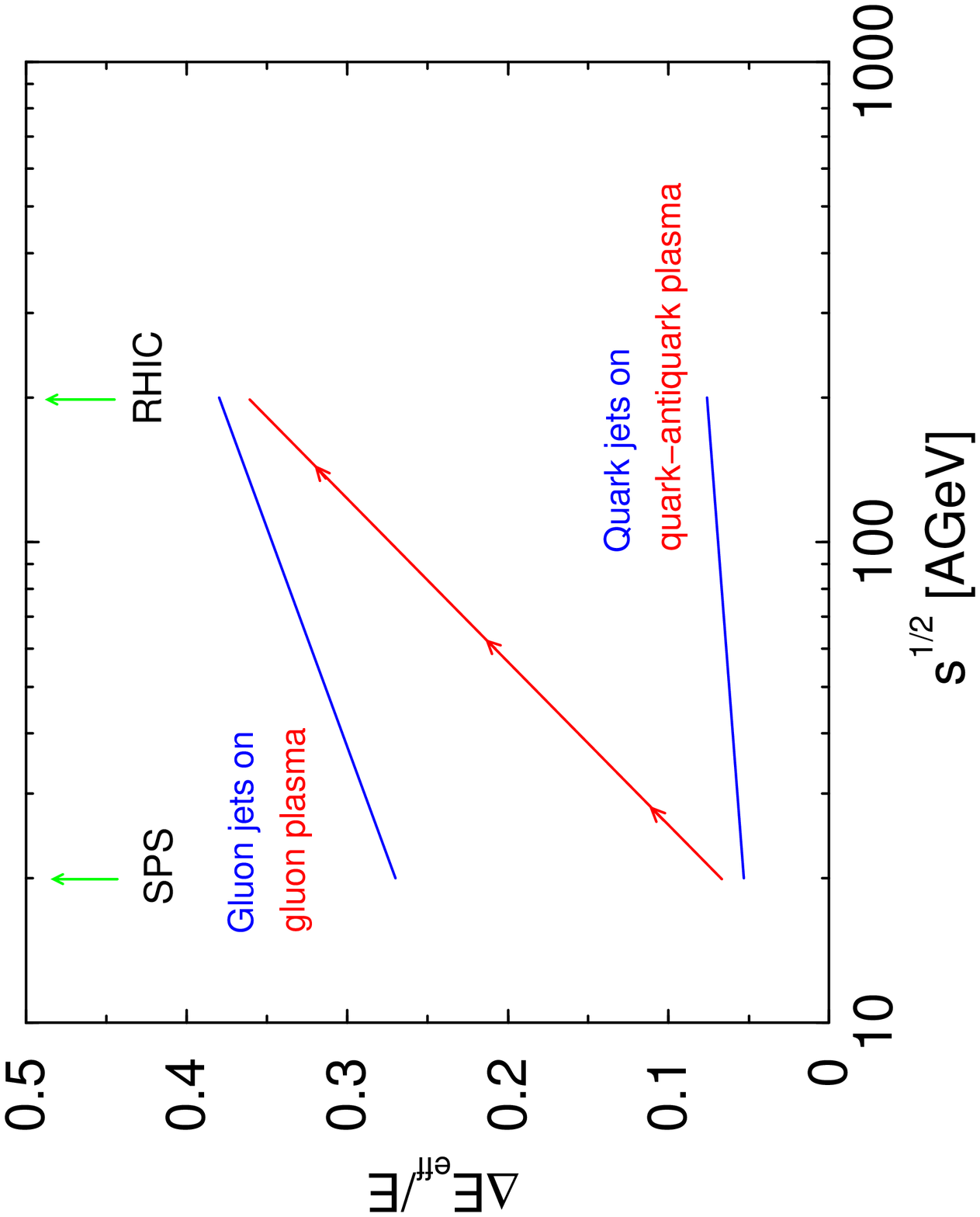}
\vspace{-4.2cm}
\end{center}
\begin{center}
\begin{minipage}[t]{11.5cm}
         {\bf Fig. 3.} {\small The {\em effective } fractional energy loss 
for $E=6$~GeV jet is plotted versus  $\sqrt{s}$  for two extreme 
scenarios.}
\end{minipage}
\end{center}

Fig.~3 illustrates the idea behind the QGP composition tomography. We contrast 
the case of gluon jets propagating through gluon plasma to the case of quarks
 propagating through quark-antiquark plasma at center of mass energies
ranging from $\sqrt{s}=20$~AGeV at SPS  to $\sqrt{s} = 200$~AGeV (the top 
RHIC energy). The variation in the effective fractional energy loss (using an
approximately constant density renormalization factor 
$Z\simeq 0.4-0.5$~\cite{FLUC}) is small $\sim 50\%$ due to the small change
of the initial density of quanta per unit rapidity (this can be most 
easily  seen from the naive picture of parton-hadron duality). In contrast, 
the changing plasma composition  and the jet representation lead to a 
factor of  $(C_A/C_F)^2 \simeq 5$ difference. While neither of the 
extreme scenarios may hold  for the  broad range of collision energies in 
question, it is possible that the partonic composition of the system  
created in heavy ion reactions  evolves as a function of $\sqrt{s}$, 
leading to pronounced differences in 
$\Delta E_{eff}(\sqrt{s})/E$ as indicated in Fig.~3.  
Faster than expected increase of energy loss with $T/T_c$ was also 
discussed in a Polyakov Loop model~\cite{Rob}.

The quenching of neutral pions is
shown in Fig.~4. In this particular example we study only the effect of
energy loss and we have not included nuclear shadowing and Cronin effect. 
We have also not included a phenomenological soft component to mock the 
part of phase space that is not perturbatively accessible and is possibly
sensitive to higher twist effects.    
In Fig.~4 the medium that drives the quenching of high $p_T$ jets at RHIC 
$\sqrt{s}=130$~AGeV is reinterpreted as composed of 
60\% gluons and 40\% quarks and antiquarks. 
At $\sqrt{s}=200$~AGeV  Au+Au  the number of quanta per unit rapidity 
 increases only by  $\sim  15\%$~\cite{PHOB}. The dashed  line 
illustrates the suppression  factor $R_{AA}$ for this slightly 
increased rapidity density of soft partons when unchanged QGP composition 
is assumed. We note that the effect is even smaller than naively 
expected ($\sim 15\%$) due to finite kinematics effects in 
computing $\Delta E/E$ and hadronization. 
Detectable increase $\sim 30 \%$ in the $\pi^0$ suppression factor  
is only possible if one assumes a transition to almost completely gluon
dominated soft background as shown in Fig.~4.

At smaller center of mass energies ($\sqrt{s}=20$~AGeV) 
we have studied the case of jets propagating through quark-antiquark
plasma. We have used  phenomenologically 
a smaller mean energy loss renormalization coefficient   
$Z=0.2$ to account for the possibly bigger energy loss fluctuations 
(more detailed studies of $Z$ are needed). 
Those effects tend to strongly reduce the effective energy loss. The 
observed suppression factor $R_{AA}$,  however, convolutes the apparent 
decrease of  $\Delta E_{eff}/E$ with the much steeper $p_T$ differential 
hadron distributions~(\ref{pqcdsp}). As a result a still significant 
suppression of neutral pions   ($R_{AA} \sim 0.5$)  remains,
 which is inconsistent with the current SPS results~\cite{WANG}.  
It is  important 
to improve the jet tomographic calculation presented in Fig.~4 to take into 
account Cronin effect and nuclear (anti)shadowing  in a more 
quantitative comparison to the WA98 data as well as to investigate the 
cause for the unusually small energy loss at SPS.

\begin{center}
\vspace*{8.7cm}
\includegraphics{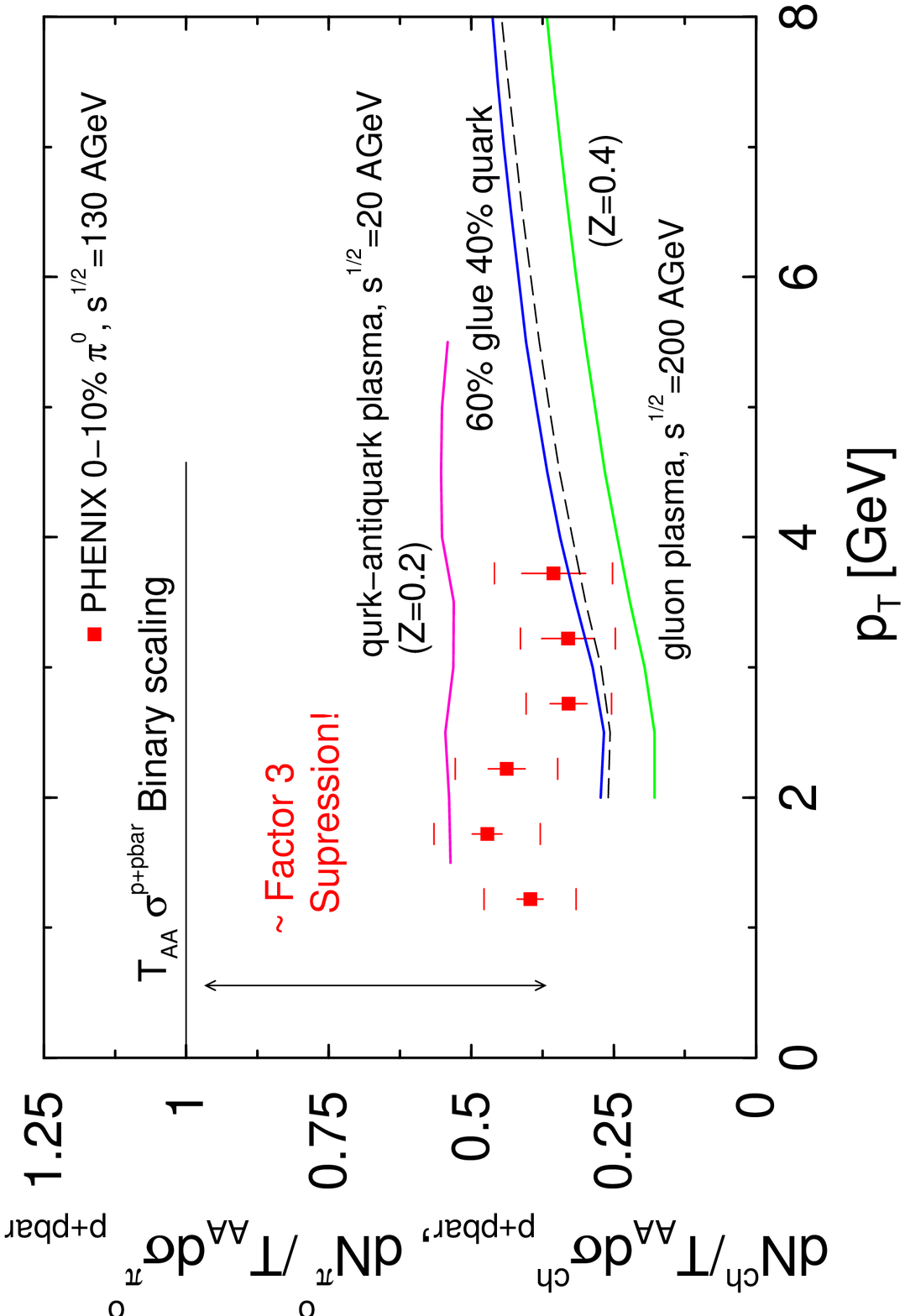}
\vspace{-3.8cm}
\end{center}
\begin{center}
\begin{minipage}[t]{11.5cm}
         {\bf Fig. 4.} {\small The $\pi^0$ suppression factor resulting from
         the non-abelian energy loss of jets is shown.}
\end{minipage}
\end{center}

\section{Conclusions}\label{concl}

We have shown that in a two component soft+hard model that incorporates jet
energy loss computed in the GLV formalism~\cite{GLV}  one can account for the
suppression of the inclusive charged hadrons and neutral pions seen 
in the $\sqrt{s}=130$~AGeV RHIC run. Jet tomography  is shown to be 
an important probe of the partonic composition of the hot and dense 
medium created in the initial stages of heavy ion reactions. In particular, 
at CERN SPS energy of $\sqrt{s}\sim 20$~AGeV large energy loss 
induced by {\em gluon} medium  on {\em gluon} jets seems  excluded by data.

\section*{Acknowledgements}

We are grateful to MTA academician I. Lovas 
for inspiring discussions on the importance
of tomographic analysis of A+A reactions, and we celebrate his 70th
birthday. Discussions with MTA septarian academicians J. Zimanyi
and J. Nemeth throughout many years are also gratefully acknowledged
on the occasion of their mutual birthdays. 
This 
trio of nuclear theorists have advanced 
nuclear physics worldwide 
and we wish them an equally productive future. 
This work was supported by the Director, Office of Science,  
Office of High Energy and Nuclear Physics, 
Division of Nuclear Physics, of the U.S. Department of Energy 
under Contract No. DE-FG02-93ER40764 and by the U.S. NSF under INT-0000211  
and  OTKA No. T029158.

\section*{Notes} 
\begin{notes} 
\item[a]  Recent reanalysis of the WA98 data  (M.M. Aggarwal {\em et al.},
Eur. Phys. J. C{\bf23},  225 (2002).) 
suggests that the enhancement
estimates should be reduced by $\sim 30 \%$. 

\item[b]  All recent approaches~\cite{BDMPS,Z,UW,GLV} that discuss the LPM
effect in QED and QCD employ the eikonal approximation together with a 
model of well separated scattering centers $\mu \lambda \gg 1$. For the case
of A+A reaction  this condition sets a natural upper limit on the opacity 
$\chi$ of a medium of finite small size $L\sim 5$~fm and is suggestive
of the dominant role of low order correlations between multiple scatterings as
demonstrated by GLV~\cite{GLV}.

\end{notes}

\vfill\eject 
\end{document}